\documentclass[letterpaper,journal]{IEEEtran}

\usepackage{amsmath,amssymb,amsfonts}
\usepackage{algorithmic}
\usepackage{array}
\usepackage[caption=false,font=normalsize,labelfont=sf,textfont=sf]{subfig}
\usepackage{textcomp}
\usepackage{stfloats}
\usepackage{verbatim}
\usepackage{graphicx}
\usepackage{cite}

\usepackage{xcolor}
\usepackage[utf8]{inputenc}
\usepackage[hyphens]{url}
\usepackage[linesnumbered, vlined,ruled]{algorithm2e}
\usepackage{multirow}
\usepackage{paralist}
\usepackage{enumitem}	
\setlist[itemize]{leftmargin=9pt,itemindent=0pt,listparindent=9pt}
\setlist[enumerate]{leftmargin=15pt,listparindent=15pt,itemindent=0pt}
\usepackage[online,flushleft]{threeparttable}
\usepackage{soul}
\usepackage{hyperref}
\usepackage{cleveref}
\usepackage{mathtools}
\usepackage{siunitx}

\crefname{formula}{formula}{formulae}

\begin{document}

\title{Abusing Commodity DRAMs in IoT Devices\\to Remotely Spy on Temperature}

\author{Florian~Frank$^*$, 
    Wenjie~Xiong$^*$, 
	Nikolaos~Athanasios~Anagnostopoulos, 
	Andr{\'e}~Schaller, 
	Tolga~Arul, 
	Farinaz~Koushanfar, 
    Stefan~Katzenbeisser, 
    Ulrich~R{\"u}hrmair,
    and~Jakub~Szefer 
\thanks{This work has been partially funded by the U.S. National Science Foundation (NSF) through grant \#1651945, by the U.S. Air Force Office of Scientific Research (AFOSR) through award FA9550-21-1-0039, by the German Research Foundation -- Deutsche Forschungsgemeinschaft (DFG), as part of the Project ``P3: Hardware-Entangled Cryptography'' (project number 236615297) of the Collaborative Research Center (CRC) 1119, as well as part of the Projects ``PUFMem'' (project number 440182124) and ``NANOSEC'' (project number 439892735) of the Priority Program ``Nano Security: From Nano-Electronics to Secure Systems'' (SPP 2253), and by the Commonwealth Cyber Initiative of the US state of Virginia.}
\thanks{$^*$ The first two authors contributed equally to this work.}%
\thanks{F. Frank, N. A. Anagnostopoulos, T. Arul and S. Katzenbeisser are with the University of Passau, Passau, Bayern, Germany. E-mails: \{florian.frank, nikolaos.anagnostopoulos, tolga.arul, stefan.katzenbeisser\}@uni-passau.de}
\thanks{W. Xiong is with Virginia Tech, Virginia, USA. E-mail: wenjiex@vt.edu}
\thanks{N. A. Anagnostopoulos and T. Arul are also with the Technical University of Darmstadt, Darmstadt, Hessen, Germany. E-mails: \{na45tisu, arul\}@rbg.informatik.tu-darmstadt.de}
\thanks{A. Schaller is with the European Organisation for the Exploitation of Meteorological Satellites (EUMETSAT), Darmstadt, Hessen, Germany. E-mail: andre@andreschaller.de}
\thanks{F. Koushanfar is with UC San Diego, California, USA. E-mail: farinaz@ucsd.edu}
\thanks{U. R{\"u}hrmair is with LMU München, München, Bayern, Germany, and with the University of Connecticut, Storrs, USA. E-mail: ruehrmair@ilo.de}
\thanks{J. Szefer is with Yale University, New Haven, Connecticut, USA. E-mail: jakub.szefer@yale.edu}
}

\markboth{Frank \MakeLowercase{\textit{et al.}}: Abusing Commodity DRAM\MakeLowercase{s} in I\MakeLowercase{o}T Devices to Remotely Spy on Temperature}%
{Frank \MakeLowercase{\textit{et al.}}: Abusing Commodity DRAM\MakeLowercase{s} in I\MakeLowercase{o}T Devices to Remotely Spy on Temperature}
\maketitle

\begin{abstract}
The ubiquity and pervasiveness of modern Internet of Things (IoT) devices opens up vast possibilities for novel applications, but simultaneously also allows spying on, and collecting data from, unsuspecting users to a previously unseen extent.
This paper details a new attack form in this vein, in which the decay properties of widespread, off-the-shelf DRAM modules are exploited to accurately sense
the temperature in the vicinity of the DRAM-carrying device. Among others, this enables adversaries to {\it remotely} and {\it purely digitally} spy on personal behavior in users' private homes, or to collect security-critical data in server farms, cloud storage centers, or commercial production lines. We demonstrate that our attack can be performed by merely compromising the software of an IoT device and does not require hardware modifications or physical access at attack time. It can achieve temperature resolutions of up to $\mathbf{0.5^\circ}$C over a range of $\mathbf{0^\circ}$C to $\mathbf{70^\circ}$C 
in practice. Perhaps most interestingly, it even works in devices that do not have a dedicated temperature sensor on board. To complete our work, we discuss practical attack scenarios as well as possible countermeasures against our temperature espionage attacks.
\end{abstract}

\begin{IEEEkeywords}
Dynamic Random Access Memory (DRAM), Internet of Things (IoT), privacy, security, temperature
\end{IEEEkeywords}

\IEEEpeerreviewmaketitle

\section{Introduction}  \label{sec:Intro}

\subsection{Motivation and Overview}

Internet of Things (IoT) devices have become more pervasive and ubiquitous than ever before in history, and are still enjoying an unbroken and continuous growth: As estimated by Taylor et al., the number of connected devices will exceed 100 billion by 2025~\cite{7365193}.
Their versatility allows applications in a large number of settings, including private and commercial uses in homes, companies, or factories.
Unfortunately, this situation also induces pressing privacy and security problems: Once an IoT device has been compromised, attackers can gather sensitive information remotely, 
as, by definition, it will be connected to the World Wide Web.

To defend against such threats, intense efforts have been made to ensure that the software of IoT devices will handle any information generated by the devices' multiple sensors in a secure manner~\cite{koeberl2014trustlite,kohnhauser2015puf,potkonjak2010trusted}. 
For example, in order to protect acoustic signals in the device environment,
 dedicated software may safeguard any information collected from the device's microphones~\cite{michalevsky2014gyrophone}.
However, even if all information collected from traditional sensors is properly protected, critical data may also be gathered from \emph{other}, unprotected device components~\cite{8383887,10.1145/3335203.3335712}, leading to {covert} and unforeseen espionage channels~\cite{10.1145/3338498.3358650,7428065,10.1145/3212480.3212489}. 
For example, while gyroscope measurements originally were considered suitable only for motion detection, researchers found ways to misuse them for measuring acoustic signals~\cite{8383887}. This allows gathering sound recordings and recognizing speech~\cite{michalevsky2014gyrophone}. 
As another example, the power usage of a mobile phone can be used to track the current position of users~\cite{michalevsky2015powerspy}, threatening their location privacy.
As long as the ability of each device component to collect critical information is not comprehensively understood, attacks of this type are hard to prevent. They put the users' privacy and security at risk, even in the presence of standard sensor-protecting measures. Consequentially, protecting single selected sensors is insufficient to guarantee a holistic protection of users.

\IEEEpubidadjcol

This article adds to this line of research. It discusses a novel method by which 
standard, widespread DRAM modules, which are part of every smart phone, laptop, or embedded device, can be abused to spy on users, and to remotely monitor the ambient temperature around the host device.
It is long known that such ambient temperature contains much security-critical information:
\begin{itemize}
    \item If measured at the victim's home, it can reveal the victim's daily routines, including holidays, or routinely repeating periods of the day during which no one is present to guard the home;
    \item If measured at a production line, it can reveal the temperature of the manufacturing process of a product;
    \item If measured at a data center, it can reveal the activity of the tenants~\cite{islam2017exploiting}.
\end{itemize}

On a technical level, our attack works by exploiting the temperature-dependent {\it ``retention times''} of DRAM cells.
The retention time of a single DRAM cell is defined as the maximal time period this cell can hold its stored value without being refreshed (i.e., how long it holds its value when the DRAM refresh operation has been disabled). It is long known that this retention time depends heavily on the ambient temperature and that it even decreases exponentially with increasing temperature~\cite{liu2013experimental,Xiong2016,schaller2017intrinsic}.
Following this observation, we turn the individual retention times of DRAM cells into a highly sensitive thermometer. In greater detail, we observe the number of flipped cells (i.e., the number of cells that change their original content) in a given DRAM memory array with disabled refresh after a certain time period has elapsed. From this number, we can then indirectly conclude the ambient temperature.
As we show in this paper, this approach allows temperature measurements over a large temperature range with a resolution of up to~\SI{0.5}{\celsius}. 

Our attack 
can be applied {\it without} measuring or registering the DRAM module under attack at multiple known temperatures in advance, as long as the general temperature-dependent characteristics of the used class of DRAM modules are known to the attacker.

It only requires compromising the software of an IoT device: Kernel access (in order to disable the DRAM refresh operation) is both necessary and sufficient for our method to work. No physical modifications such as hardware Trojans or the like are required, and neither is physical access to the device at attack time. Finally, and perhaps most interestingly, our technique can even be used to remotely spy on the temperature in devices that {do not} contain temperature sensors~{\it at~all}.

\subsection{Our Contributions}

This article is an extended version of our earlier conference paper~\cite{8714882}. In the original work, it was demonstrated that DRAM decay can be used to measure the ambient temperature, only utilizing modified software on IoT devices. An attacker can practically conduct the DRAM decay enrollments at a constant ambient temperature to later map the DRAM decay measurement results to different temperatures and guess the user's behavior or environmental changes. The temperature resolution was shown to be as good as \SI{0.5}{\celsius} in commodity, off-the-shelf IoT devices, enabling attackers to measure fine-grained temperature changes around the IoT devices. This extended version now contains the following additional contributions: 

\begin{itemize}
    \item Additional measurements were carried out with higher precision and using at a greater temperature range compared with the original paper. Now, each measurement is taken within a stable temperature environment (a temperature chamber) with a deviation of at most \SI{0.1}{\celsius}, and over a temperature range from \SI{0}{\celsius} to \SI{70}{\celsius} in \SI{2.5}{\celsius} increments. In the original paper, only measurements from \SI{20}{\celsius} to \SI{45}{\celsius} were~taken.
    
    \item Additionally, the dependency between bit-flips and the ambient temperature is now described more precisely using these new measurements. A new approximation function was developed to calculate the ambient temperature without capturing enrollment measurements over the whole temperature range on the spying board. Instead, the measurements are taken from a similar device, and the temperature to bit-flips characteristics of that similar board are used to spy the temperature on the spying board using only a single enrollment at a known temperature from this board.
    This is possible with an accuracy of less than \SI{1}{\celsius}, requiring only one enrollment measurement on the spying board. If a board is used for which enrollment measurements over the whole temperature range are available, a precision higher than \SI{0.5}{\celsius} can now be achieved.
    
    \item Subsequently, attacks are demonstrated in two scenarios. In the first one, the workload of a server is approximated by measuring the temperature in its vicinity, only using the bit-flips in a DRAM region. In the second scenario, a temperature spying attack is demonstrated in an IoT~environment, e.g., in the context of a smart home.
    
    \item Finally, this extended version now demonstrates the first countermeasures to prevent such kinds of attacks in practice. For example, we experimentally prove that putting the device inside a closed box will consequently distort the temperature measurements, disabling the attacks.
\end{itemize}
\subsection{Related Work}

Besides the original paper~\cite{8714882}, various related works exist that describe the dependence of the DRAM retention time on ambient temperatures, often in relation to DRAM Physical Unclonable Functions (PUFs). 

Two well-known works that investigate the DRAM retention effect under different temperatures in a PUF context, were published by Anagnostopoulos et al.~\cite{8635789} and Schaller et al.~\cite{schaller2018decay}. In these publications, row-hammering PUFs and DRAM-retention PUFs were examined on Panda\-Boards and Intel Galileo boards. 
They describe the dependence of DRAM retention PUFs on the supply voltage and ambient temperature, tested from temperatures ranging from \SI{25}{\celsius} to \SI{40}{\celsius} ~\cite{8635789} and from \SI{40}{\celsius} to \SI{80}{\celsius}~\cite{schaller2018decay}.

Another work describing the dependence of the DRAM retention time on the temperature was published by Wang et al.~\cite{8388826}. This paper focuses on computing at cryogenic temperatures, for example, required to implement quantum computers. For this reason, the retention of DRAM memory modules was examined on temperatures from \SI{358}{\kelvin}, \SI{77}{\kelvin} and \SI{263}{\kelvin}. Müelich et al.~\cite{9003355} present a theoretical model which captures the instabilities of PUF responses. DRAM retention PUFs are evaluated on temperatures from \SI{25}{\celsius} to \SI{90}{\celsius}.

The existing works that come closest to our work include Tian et al.~\cite{10.1145/3373087.3375322}, who present techniques on how to fingerprint FPGAs in cloud infrastructures. One contribution of this paper is to monitor the temperature in the vicinity of FPGAs using bit-flips resulting from DRAM decay.
Another related paper about information leaks in cloud infrastructures was published by Giechaskiel et al.~\cite{giechaskiel2021cross}. Here bit-flips in DRAM memory modules of FPGAs were used to monitor the temperature in the vicinity of FPGAs within data centers.
A second paper published by Giechaskiel et al.\cite{10.1145/3534972} describes the same attack as in the previous paper in more detail and additionally demonstrates an attack using 24 FPGAs to monitor the temperature over 24 hours. 

In comparison to these two latter works, our paper provides a more detailed evaluation of various temperatures; two different attacks are demonstrated; and countermeasures are implemented and proven. Furthermore, as already mentioned, this work is a journal version of the original article~\cite{8714882}, which appeared earlier than the above-mentioned publications ~\cite{10.1145/3373087.3375322} and \cite{giechaskiel2021cross}.

\subsection{Organization of This Paper}

The rest of this paper is organized as follows: Section~\ref{sec:Background} presents background information on DRAMs and their decay behavior when the refresh operation is disabled. Section~\ref{sec:BasicSteps} discusses the basic steps in our temperature spy attacks, including the experimental setup, the role of indicator cells for one attack scenario, and an approximation function, describing the dependency between bit-flips and temperature, used in the second type of attack. Additionally, information is provided on how the DRAM cells can be accessed during runtime.
The subsequent Section~\ref{sec:Evaluation} evaluates the methodologies introduced in the previous section and presents as well as analyses the measurement results. Countermeasures are suggested, implemented, and successfully demonstrated in Section~\ref{sec:Countermeasures}. Finally, Section~\ref{sec:Conclusions} concludes this work.
\section{Background: DRAM Cells and Their\break Decay Characteristics} \label{sec:Background}

\begin{figure*}[!b]
	\centering
	\includegraphics{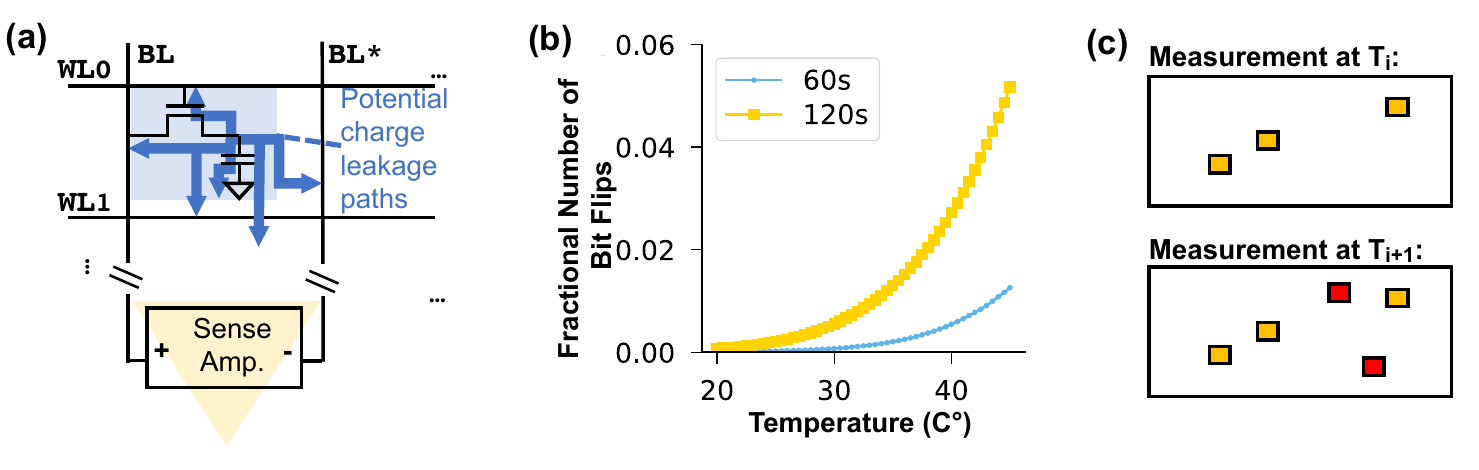}
	\caption{(a) DRAM cell schematic. (b) Temperature dependency of the fractional bit-flips for DRAM modules from a tested Intel Galileo board. (c) Illustration of a sample DRAM array decay measurement at temperatures $T_i$ and $T_{i+1}$, both with the same decay time $t$; the highlighted cells are the cells where a bit-flip occurs -- for the same decay time, more cells flip at the higher temperature (illustrated as red cells). }
	\label{fig:DRAM_cell}
\end{figure*}

DRAM is one of the most widely used memory types in computer devices.
In DRAM, each bit of data is stored in a DRAM cell that consists of an access transistor and a capacitor, as shown in \Cref{fig:DRAM_cell}~(a).

Each capacitor of the DRAM cell has two states, charged and discharged, which are used to store one bit of data.
Word Lines (WL) are used to enable the access transistors, and Bit Lines (BL) to read out data.
Data from two complementary bit lines (BL and BL*) are amplified through a sense amplifier.
The amplifier is used to convert analog voltage levels into either full $V_{DD}$ or \SI{0}{\V} depending on the present voltage on the capacitor.

DRAM is a volatile memory, and the stored data will be lost if refresh or power is turned off.
To prevent data loss, each DRAM cell requires a periodic refresh. Most memory modules have a refresh period of \SI{64}{\milli\second}.
This is because capacitors lose charge over time.
\Cref{fig:DRAM_cell}~(a) shows possible paths through which the charge on each capacitor can leak.

When DRAM refresh is disabled, charged cells will steadily lose charge.
If a charged cell loses enough charge, it becomes discharged, and the stored bit of data flips.
The loss of charge over time is referred to as {\em DRAM decay}.
The time, a cell can keep a bit value without refresh is called the {\em retention time}.
Different DRAM cells have different retention times.
Thus, if the refresh operation is disabled for a longer time, more cells' retention times are exceeded, and more bit-flips appear.
Meanwhile, if a cell is initialized to a discharged state, its value will never flip as there is no charge to leak.

A DRAM decay measurement is a measurement showing which DRAM cells have flipped in a DRAM region after a given decay time $t$ has elapsed.
To perform a DRAM decay measurement, first, a DRAM region is selected and the cells in this region are initialized to a known value.
For example, this work uses logical \texttt{0} as the initial value of all the cells\footnote{Note, that some DRAM cells map logical \texttt{0} to the charged
state, whilst others map logical \texttt{1} to the charged state.
The exact mapping is not published by DRAM manufacturers, 
but we have empirically derived that about half the cells in the tested DRAM modules map logical \texttt{0} to the charged state.
The cells that map logical \texttt{0} to the discharged state simply do not contribute any value to the measurement, but also do not interfere with it.}.
The DRAM region is then allowed to decay, i.e., the refresh operation is disabled, for time $t$.
After the elapsed time $t$, the DRAM region is read, to observe which cells have flipped their initialized logical value of \texttt{0} to logical \texttt{1}. These cells are the ones that have decayed during time~$t$.
Moreover, in an independent field of study, DRAM decay is leveraged for creating a PUF, which can be used as security primitive for authentication and key storage, 
e.g.,~\cite{Xiong2016,schaller2017intrinsic,8635789,schaller2018decay,rosenblatt2013field,rosenblatt2013self,rahmati2015probable,sutar2016d,mexis2021lightweight,8771257}. For instance, PUFatt~\cite{kong2014pufatt} demonstrates an ALU PUF-based secure remote attestation scheme for embedded systems. SHAIP~\cite{hussain2018shaip} is a PUF-based mutual authentication framework with unlimited number of authentications and privacy-preserving property. BIST-PUF~\cite{hussain2014bist} enables real-time assessment of PUF’s unpredictability and stability in hardware. The emerging trends and challenges of PUFs and robust protocols are discussed in ~\cite{rostami2014quo, rostami2014robust}. Similar to other PUF solutions \cite{pappu2002physical,gassend2002silicon,lugli2013physical,csaba2010application,nguyen2018interpose,ruhrmair2014pufs}, these PUFs may potentially provide
improved resilience against invasive \cite{nedospasov2013invasive} and side-channel attacks~\cite{ruhrmair2013power}.

The said DRAM decay, among other things, depends highly sensitively on the temperature~\cite{liu2013experimental,Xiong2016,schaller2017intrinsic,8635789,8771257},
with a higher temperature accelerating the charge leakage and the decay process. 
The fractional number of bit-flips (i.e., the number of bit-flips in the DRAM region divided by the size of this region)
for our Intel Galileo board~\cite{GalileoGen2} is shown in \Cref{fig:DRAM_cell}~(b).
The number of bit-flips increases as the temperature increases.
Also, using a decay time of~$t=\SI{120}{\second}$ will result in more bit-flips than using a shorter decay time of~$t=\SI{60}{\second}$, for example.

\Cref{fig:DRAM_cell}~(c) illustrates example DRAM decay results for the same decay time $t$, but at temperatures $T_{i}$ and $T_{i+1}$ ($T_{i+1} > T_i$).
More bit-flips appear at the higher temperature $T_{i+1}$, compared to the lower temperature $T_i$. 
Also, bit-flips that occur at a lower temperature are a subset of the bit-flips that occur at a higher~temperature.
\section{Basic Steps in Our Temperature Spy Attack}  
\label{sec:BasicSteps}
Many IoT devices may not have a dedicated temperature sensor -- this work shows that even in absence of a temperature sensor, attackers can still leverage DRAM cells in IoT devices to obtain the ambient temperature.
The attacker first needs to compromise the remote IoT device to be able to control the DRAM refresh. Usually, he or she needs to compromise the kernel (for some devices also the firmware needs to be modified) to measure the DRAM decay.
Many IoT devices are vulnerable to exploits that can give kernel privileges.
Once the device is compromised, the attacker needs to take $m$ enrollment measurements.
We show that the measurements can be taken using different decay times at a constant, but possibly unknown, temperature.
Based on the enrollments, the attacker can map the DRAM decay results to temperature.
Thus, the attacker simply performs one DRAM decay measurement and acquires the temperature.

\subsection{Enrolling DRAM Decay}

To map DRAM decay measurements of a DRAM region to a particular DRAM module, a so-called {\em enrollment} is executed where multiple measurements are taken.
Normally, the enrollment should be performed at a fixed decay time $t$ and $m$ different temperatures $\{T_0, T_1, T_2, ..., T_m\}$,
covering the temperature range of~interest.

Often it is not possible for an attacker to control the ambient~temperature during the enrollment.
This is the reason why this work discusses the precision of the temperature spying attack, when the attacker can execute an enrollment with $m$ different temperatures, as well as when only one constant temperature measurement can be used for the enrollment.

It is possible to use only one temperature because the DRAM decay at different temperatures can be simulated by measurements with multiple decay times at a constant temperature, for example, the decay result of decay time $2t$ at enrollment temperature $T$ can be approximated by using the decay result of decay time $t$ at temperature $T{+}\SI{10}{\celsius}$. The relationship between temperature and time is further derived and evaluated in \Cref{sec:const_enroll}.
In this way, the attacker takes measurements at a constant temperature $T_0$ 
for decay times $\{t_0, t_1, t_2, ..., t_m\}$ to simulate decay results at $\{T_0, T_1, T_2, ..., T_m\}$ 
for decay time $t_0$ at each of these temperatures. In \Cref{sec:const_enroll}, we experimentally validate this~approach.

\subsection{Mapping DRAM Decay to Temperature Changes}

Given the $m$ enrollment measurements simulating different temperatures, a mapping between the DRAM decay results and the ambient temperatures can be generated by counting the number of bit-flips in the $i$-th enrollment and mapping that number of bit-flips to temperature $T_i$.
Later, given a DRAM decay measurement, the number of bit-flips can be counted and compared with the enrollment measurements.
The temperature of the measurement is seen to be the same as the temperature of the enrollment measurement with the most similar number of bit-flips. However, when counting the bit-flips, the whole DRAM region measured needs to be read. This introduces memory bandwidth and computational overhead during measurement time.

To overcome this challenge, we propose two different approaches: (1)~For some use cases, it is sufficient to calculate the number of bit-flips on the attacked device and only send this number to the attacker, which allows saving a huge amount of network bandwidth, and also decreases the probability that the attack is noticed. 
(2)~Some other use cases require more information. That's when {\em indicator cells} are used. By using only a subset of the cells in the DRAM region, the memory bandwidth and computational overhead is reduced.
With enrollment measurements simulating temperatures $T_{i}$ and $T_{i+1}$ ($T_{i+1} > T_{i}$), to choose the indicator cells for $T_{i}$, the two enrollments are compared, 
and cells that flip at $T_{i+1}$ but not at $T_{i}$ are the {\em candidate indicator cells}. For example, candidate indicator cells are highlighted in red in \Cref{fig:DRAM_cell}~(c). 
Among the candidate cells, $l$ cells are selected as \emph{indicator cells}. Depending on the expected noise level 
(see \Cref{sec:eval}) the number $l$ can be increased.
Typically, an odd number of cells is needed to allow majority voting.
If $l$ candidate indicator cells cannot be identified, a larger DRAM region or a longer decay time $t$ should be used.
The locations of all the indicator cells for all temperatures need to be saved.
Later, during a temperature spy attack, only the $l * (m-1)$ indicator cells (indicator cells for all enrolled temperatures)
need to be read. 
For each potential temperature, the majority vote of $l$ indicator cells is used to decide whether the current temperature is above $T_{i}$.
Then, after at most $m-1$ majority votes, the current temperature $T_{cur}$ is known.
This can save memory bandwidth of several orders of magnitude because only dozens of indicator cells need to be stored instead of all the KiBs or MiBs of cells in the DRAM region being measured at measurement time.

\subsection{DRAM Decay Measurement at System Runtime}

It is not trivial to make the DRAM decay measurement at system runtime without hardware changes, because it is not possible to disable the DRAM refresh for arbitrary DRAM regions. 
If the whole DRAM module's refresh is disabled, all content of the memory will eventually decay and errors in the memory contents will cause the system to crash.
As a solution, similar to the approach of~\cite{Xiong2016}, this work uses a kernel module to disable the refresh of the whole DRAM module
while issuing extra memory accesses to the memory regions holding the critical system data.
Each DRAM access also behaves as a refresh, so the system data that are explicitly accessed will not decay.
At the same time, the other cells in the DRAM, which are not accessed, will decay.

\section{Experimental Evaluation of Our Attack}  \label{sec:Evaluation}
\label{sec:eval}

In our evaluation, first, we describe the setup of our experiments and how the tests are executed.
The next section, \Cref{sec:DRAM_temp_sensor}, investigates the resolution of DRAM as a temperature sensor.
Then, \Cref{sec:const_enroll} shows that it is possible to take enrollment measurements at a constant temperature. 
In the subsequent section \ref{sec:approximation_function}, a new approximation function is introduced, which allows spying on devices using the approximated DRAM decay characteristic of a different device.
In \Cref{sec:attack_demo} and \Cref{sec:attack_comp}, attack examples are presented and the complexity of the attack is discussed. 

\subsection{Experimental Setup and Test Execution}
\label{sec:Experimental Setup and Test Execution}

The evaluation of the number of available indicator cells and their reliability is conducted on Intel Galileo Gen~2~\cite{GalileoGen2} IoT development boards, which are equipped with an Intel Quark X1000 SoC and with two $128$MiB DDR3-SDRAM modules from Micron.
There, the DRAM decay measurements can be executed by a modified firmware, or by loading a kernel module to measure the DRAM decay in the chosen DRAM region during operation. 
In total, four Intel Galileo boards are measured.
To allow for precise evaluation of the temperature-dependent characteristics of DRAM modules, a TestEquity 1007C~\cite{thermalchamber} and a Weisstechnik LabEvent thermal chamber~\cite{weisstechnik} are used to control the ambient~temperature. 

First, DRAM decay measurements are performed at \mbox{$\mathcal{T}=\{20, 21, ..., 45\}$\si{\celsius},} where $dT=T_{i+1}-T_{i}$ denotes the step between temperature points in $\mathcal{T}$, thus here $dT=1$\si{\celsius}.
\Cref{fig:bitflip_size} shows the number of candidate indicator cells , i.e., bit-flips at temperature $T_{i+1}$ but not $T_{i}$.
The results are the average of DRAM regions on four Galileo boards.
Two decay times of $t=\SI{60}{\second}$ and $t=\SI{120}{\second}$ are tested, with DRAM region sizes of $512$KiB, $1$MiB, and $2$MiB.

The number of candidate indicator cells depends on the DRAM region size and the decay time $t$.
Within the range, the smallest number of candidate indicator cells occurs when the attacker measures a $512$KiB DRAM region at \SI{20}{\celsius} -- one of the tested boards gives only 2 indicator cells.
However, $1$MiB DRAM is sufficient to support $l=3$ or $l=5$, and thus supports error rates of up to $40\%$.

In this extended version, additional measurements with a temperature range of $\mathcal{T}_{new} = \{0, 2.5, ..., 70\}$\si{\celsius} are taken, to examine especially lower and higher temperatures. Here, we notice that a larger step size of $dT_{new}=\SI{2.5}{\celsius}$ is sufficient to approximate the temperatures between the \SI{2.5}{\celsius} steps. These temperatures are evaluated with further decay times of $t=\SI{60}{\second}$, $t=\SI{120}{\second}$, $t=\SI{180}{\second}$ and $t=\SI{240}{\second}$, and a memory region size of 1 MiB.
These measurements are visualized in Figure \ref{fig:approximation} (a) to (d). There, the total amount of bit-flips is measured per temperature and decay time. To increase the precision of the measurements, first, the firmware of the Galileo boards is modified in a way such that, after the decay time elapsed, the reserved memory region is immediately copied to a different region in the RAM and the refresh is reactivated. This avoids additional bit-flips during reading and, thus, avoids distortion of the measurement results. Additionally, a Raspberry Pi controls the climate chamber and the test execution on the Galileo boards.
It starts the tests on the Galileo boards and simultaneously monitors as well as controls the temperature of the climate chamber, which requires some time until the temperature stabilizes. That is the reason why after reaching the target temperature, the test program on the Raspberry Pi checks for \SI{90}{\second} if the temperature is within the acceptance range, which is in our case \SI{0.1}{\celsius}. Only if this is the case, the experiments are started.

\subsection{DRAM Temperature Sensor Based on Indicator Cells}
\label{sec:DRAM_temp_sensor}
To show that DRAM decay can be used to observe the device's ambient temperature, this section answers the following~questions:
\begin{enumerate}
	\item What level of error rate in the measurements can be corrected by the majority voting of $l$ indicator cells?
	\item How many candidate indicator cells are available for a given DRAM region size in the tested devices?
	\item Are the chosen indicator cells reliable?
	\item How sensitive is DRAM decay to temperature changes?
\end{enumerate}

\begin{figure*}[t]
    \centering
    \begin{minipage}{0.31975\linewidth}
        \centering
        \includegraphics[width=\linewidth]{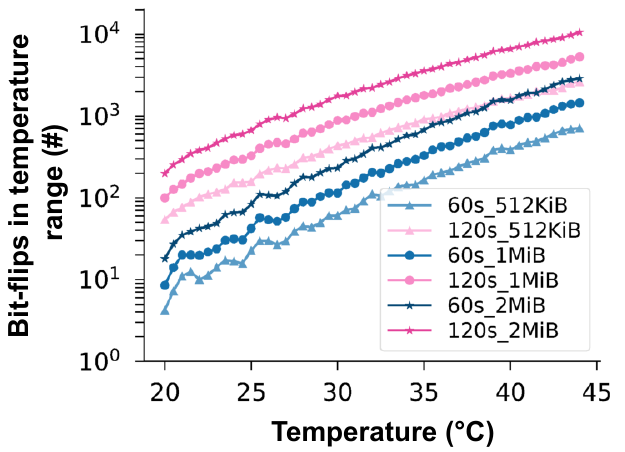}
        \caption{Number of bit-flips in the temperature range $[T,T{+}1\si{\celsius}]$ versus temperature $T$ for different DRAM region sizes.}
        \label{fig:bitflip_size}
    \end{minipage}
    \hfill
    \begin{minipage}{0.31975\linewidth}
        \centering
        \includegraphics[width=0.9\linewidth]{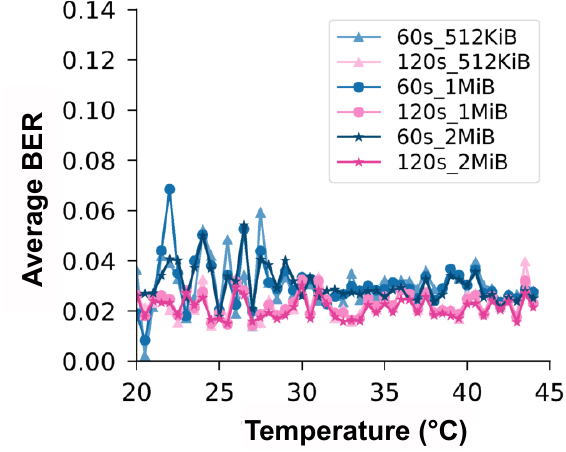}
        \caption{Average BER in the temperature range $[T,T{+}1\si{\celsius}]$ versus temperature $T$ for different DRAM region sizes.}
        \label{fig:BER_size}
    \end{minipage}
    \hfill
    \begin{minipage}{0.31975\linewidth} 
        \centering
        \includegraphics[width=0.9\linewidth]{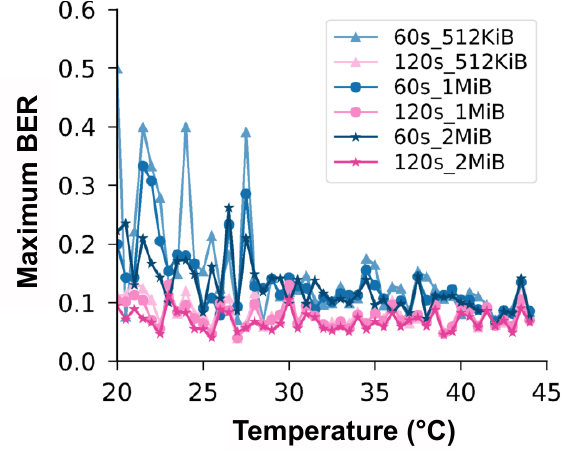}
        \caption{Maximum BER in the temperature range $[T,T{+}1\si{\celsius}]$ versus temperature $T$ for different DRAM region sizes.}
        \label{fig:max_BER_size}
    \end{minipage}
\end{figure*}

\begin{figure*}[t]
    \centering
    \begin{minipage}{0.31975\linewidth}
        \centering
        \includegraphics[width=\linewidth]{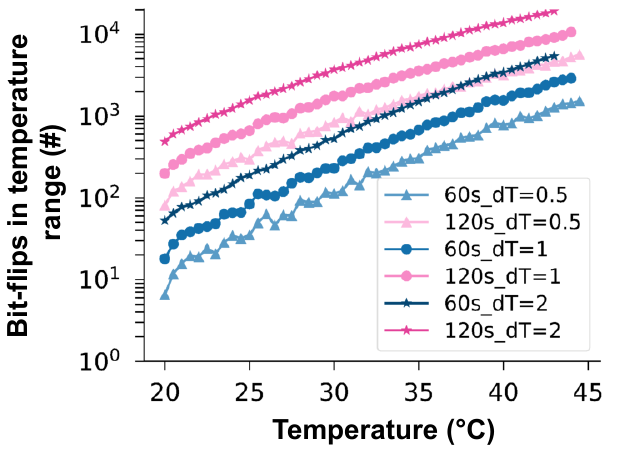}
        \caption{Number of bit-flips in the temperature range $[T,T{+}dT]$ versus temperature $T$ for different $dT$ values in a $2$MiB DRAM region.}
        \label{fig:bitflip_dT}
    \end{minipage}
    \hfill
    \begin{minipage}{0.31975\linewidth}
        \centering
        \includegraphics[width=0.9\linewidth]{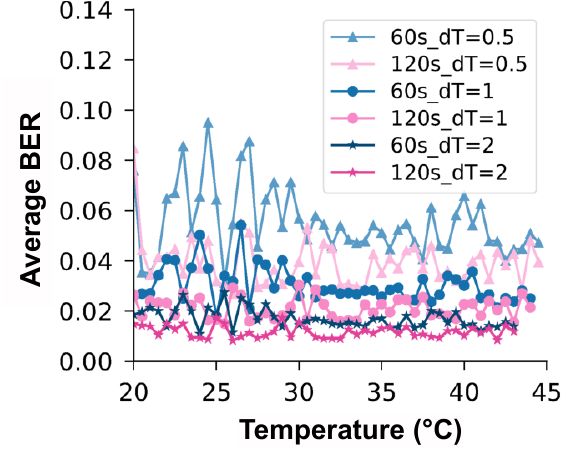}
        \caption{Average BER in the temperature range $[T,T{+}dT]$ versus temperature $T$ for different $dT$ values in a $2$MiB DRAM region.}
        \label{fig:BER_dT}
    \end{minipage}
    \hfill
    \begin{minipage}{0.31975\linewidth}
        \centering
        \includegraphics[width=0.9\linewidth]{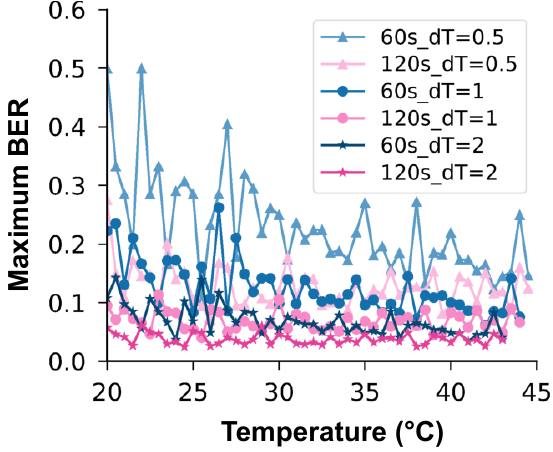}
        \caption{Maximum BER in the temperature range $[T,T{+}dT]$ versus temperature $T$ for different $dT$ values in a $2$MiB DRAM region.}
        \label{fig:max_BER_dT}
    \end{minipage}
\end{figure*}

\medskip{\noindent\textbf{Supported Measurement Error Rates}}.
When the attacker attempts to derive the temperature from the DRAM decay measurements, he or she will use $l$ indicator cells
and perform majority votes. The minimum value of $l$ is $3$. With $l=3$, an error rate of up to $33\%$ can be corrected
by the majority vote.
With $l=5$, the majority vote can correct an error rate of $40\%$, and so forth. In practice, the attacker can choose $l$ based on the noise and the number of available candidate indicator cells in the DRAM region.

\medskip{\noindent\textbf{Reliability of Indicator Cells}}.
An ideal indicator cell for $T_{i}$ should never flip at $T_{i}$ and always flip at $T_{i+1}$. 
To evaluate the reliability, at each temperature five measurements are taken for each of the four Galileo boards. 
The first measurement is used as enrollment, and the other four spy measurements are used for testing the reliability.
\Cref{fig:BER_size,fig:max_BER_size} show the average and maximum Bit Error Rate (BER) of the spy measurements. 
The BER for each temperature $T_i$ is calculated as follows: 
\begin{enumerate}
	\item in the enrollment measurement, the number of candidate indicator cells are counted; 
	\item in the spy measurement, if an indicator cell flips at $T_{i}$ or an indicator cell does not flip at $T_{i+1}$, this cell is seen as an error;
	\item  the number of errors is counted and divided by the result from step (i) to compute the BER. 
\end{enumerate}

\Cref{fig:BER_size,fig:max_BER_size} show the average and maximum BER across the four boards and four spy measurements.
As shown in \Cref{fig:BER_size,fig:max_BER_size}, a longer decay time, a larger DRAM region size, or a higher temperature will result in a smaller BER. This is because a longer decay time, a larger DRAM region, and a higher temperature, each yield more indicator cells (and more reliable indicator cells).
The average BER is much smaller than the maximum case, meaning that there are only a few cases where the BER is high. 
As shown in \Cref{fig:max_BER_size}, to obtain reliable results, at least $1$MiB DRAM region size is needed to achieve a BER of less than $33\%$ for $l=3$.

\medskip{\noindent\textbf{Temperature Sensitivity}}.
We further explore different temperature resolutions, where $dT=0.5$\si{\celsius}, $1$\si{\celsius}, $2$\si{\celsius} separately. 
The results in \Cref{fig:bitflip_dT,fig:BER_dT,fig:max_BER_dT} are retrieved from all four Galileo boards, with a DRAM region of $2$MiB and a decay time of either \SI{60}{\second} or \SI{120}{\second}.
The data are processed in the same way as in \Cref{fig:bitflip_size,fig:BER_size,fig:max_BER_size}.
As shown in \Cref{fig:bitflip_dT}, at least 5 indicator cells can be found in $2$MiB regions in the temperature range \mbox{[$T_i$, $T_i+dT$]} for all $T_i$ and $dT$ considered. 

The average BER, shown in \Cref{fig:BER_dT}, is again much smaller than the maximum case, however \Cref{fig:max_BER_dT} shows that when $t=\SI{120}{\second}$ and $dT=\SI{0.5}{\celsius}$, 
the maximum BER can still be corrected by the majority vote of $l=3$ cells. Nevertheless, with a decay time $t=\SI{60}{\second}$ and $dT=\SI{0.5}{\celsius}$, 
the maximum BER is too large to be corrected by the majority vote of $l=3$ cells and, thus, a larger DRAM region or longer decay time should be used. 

An additional approach is discussed in this extended version. Here, ten measurements are captured per temperature and decay time and the average amount of bit-flips of these measurements are used to approximate the temperature. With this simple method, the same precision can be achieved as in the previous approach.

\subsection{Enrollments at a Constant Temperature}
\label{sec:const_enroll}

In a further investigation, we consider how the attacker can take enrollments at a fixed temperature with different decay times and derive the expected DRAM decay at other temperatures, even if he or she never enrolled the device at these temperatures.
In particular, we show how the DRAM decay with decay time $t_{real}$ and temperature $T_{real}$ can be 
simulated by a measurement with decay time $t_{sim}$ and temperature $T_{sim}$. 

We denote $\Delta T_{rs} = T_{real}-T_{sim}$, which is the temperature difference between the real temperature ($T_{real}$),
and the temperature that the attacker can measure during enrollment to simulate the real temperature ($T_{sim}$).
As indicated in the works of Xiong et al.~\cite{Xiong2016} and Schaller et al.~\cite{schaller2018decay}, the DRAM decay time $t_{{sim}}$ and $t_{real}$, and temperature $\Delta T_{rs}$ have the following relationship:
\begin{equation}
    t_{sim}= t_{real} * e^{k\Delta T_{rs}}.
    \label{equ:temp}
\end{equation}

Furthermore, we test that identical models of DRAM chips have the same temperature index $k$, 
so the attacker can compute $k$ using his or her own device (where he or she can control the temperature), to then use that $k$ for the attack on a remote device.

To validate \Cref{equ:temp}, the simulation measurements are taken at $T_{sim}=\SI{25}{\celsius}$ and $T_{sim}=\SI{30}{\celsius}$
to simulate the DRAM decay at the temperature ranges $T_{real}=$\SI{20}{\celsius} to \SI{40}{\celsius} and $T_{real}=$\SI{25}{\celsius} to \SI{45}{\celsius}, respectively. 
The measurements are designed to simulate the decay time of both \SI{60}{\second} and \SI{120}{\second}. 
To simulate decay time $t_{real}=\SI{60}{\second}$ (\SI{120}{\second}), ten different decay times in the range of $t_{sim}=\SI{45}{\second}$ to \SI{160}{\second} (\SI{90}{\second} to \SI{320}{\second}) are~measured.

To estimate the temperature index $k$, the simulation measurements and the real measurements are compared. 
For each simulation measurement, we find the real measurement that has the most similar number of bit-flips, and record the pair $t_{sim}$ and $\Delta T_{rs}'$. 
With the pairs of $t_{sim}$ and $\Delta T_{rs}'$ across $t_{real}=\SI{60}{\second}$ or \SI{120}{\second}, $T_{sim}=$\SI{25}{\celsius} or \SI{30}{\celsius}, the relevant $T_{real}$ ranges, and four Intel Galileo boards, the best-fit temperature index of $k=0.07$ can be computed.

To show that the simulation measurements are similar to the real measurements, using $k$, we compute the $\Delta T_{rs}$ for each $t_{sim}$.
We then compare each pair of real measurement ($t_{real}$, $T_{real}$) and simulation measurements ($t_{sim}$, $T_{sim}$).
To compare the two measurements, we use the {\em Jaccard Index}, which is a well-established metric to compare the similarity of two different data sets and thus very suitable for this use case.~\cite{jaccard1901etude}.
Let $\mathcal{R}$ and $\mathcal{S}$ denote the set of bit-flips in the real measurement and the simulation 
measurement, respectively. The Jaccard Index is calculated by 
$J=\frac{\lvert\mathcal{S}\cap \mathcal{R}\rvert}{\lvert\mathcal{S} \cup \mathcal{R}\rvert}$.
If the resulting $J$ is close to 1, it indicates high similarity between the two measurements.

\begin{figure}[t]

    \centering
    \begin{minipage}[t]{0.475\textwidth}
    \includegraphics[width=\linewidth]{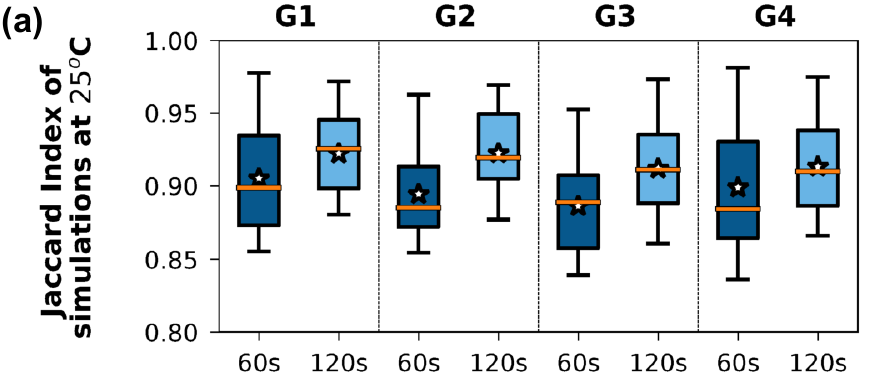}
    \end{minipage}
    \begin{minipage}[t]{0.475\textwidth}
    \includegraphics[width=\linewidth]{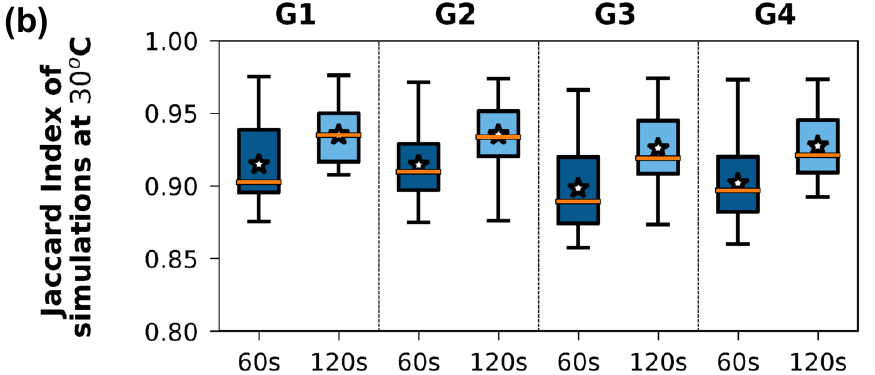}
    \end{minipage}
    \caption{Jaccard Index between simulation measurements at (a)~\SI{25}{\celsius} or (b)~\SI{30}{\celsius} and real measurements with a decay time of \SI{60}{\second} or \SI{120}{\second}.}
    \label{fig:30_enroll}
    \label{fig:25_enroll}
\end{figure}

\Cref{fig:25_enroll} shows the distribution of the Jaccard Index for $t_{real}=$ \SI{60}{\second} or \SI{120}{\second} for all four Intel Galileo boards. 
Each box contains ten different simulation decay times and the corresponding real temperatures. 
The stars indicate the average of the data set, the orange bars indicate the median. 
The Jaccard Index is higher than $0.85$ in almost every case, indicating that the simulation measurements and the real measurements are very similar. 
Thus, the enrollments can be taken at a fixed temperature to cover a range of different~temperatures.

\subsection{Approximation of the Bit-Flip Temperature Dependency}
\label{sec:approximation_function}

\begin{figure*}[!ht]
    \includegraphics[width=0.95\linewidth]{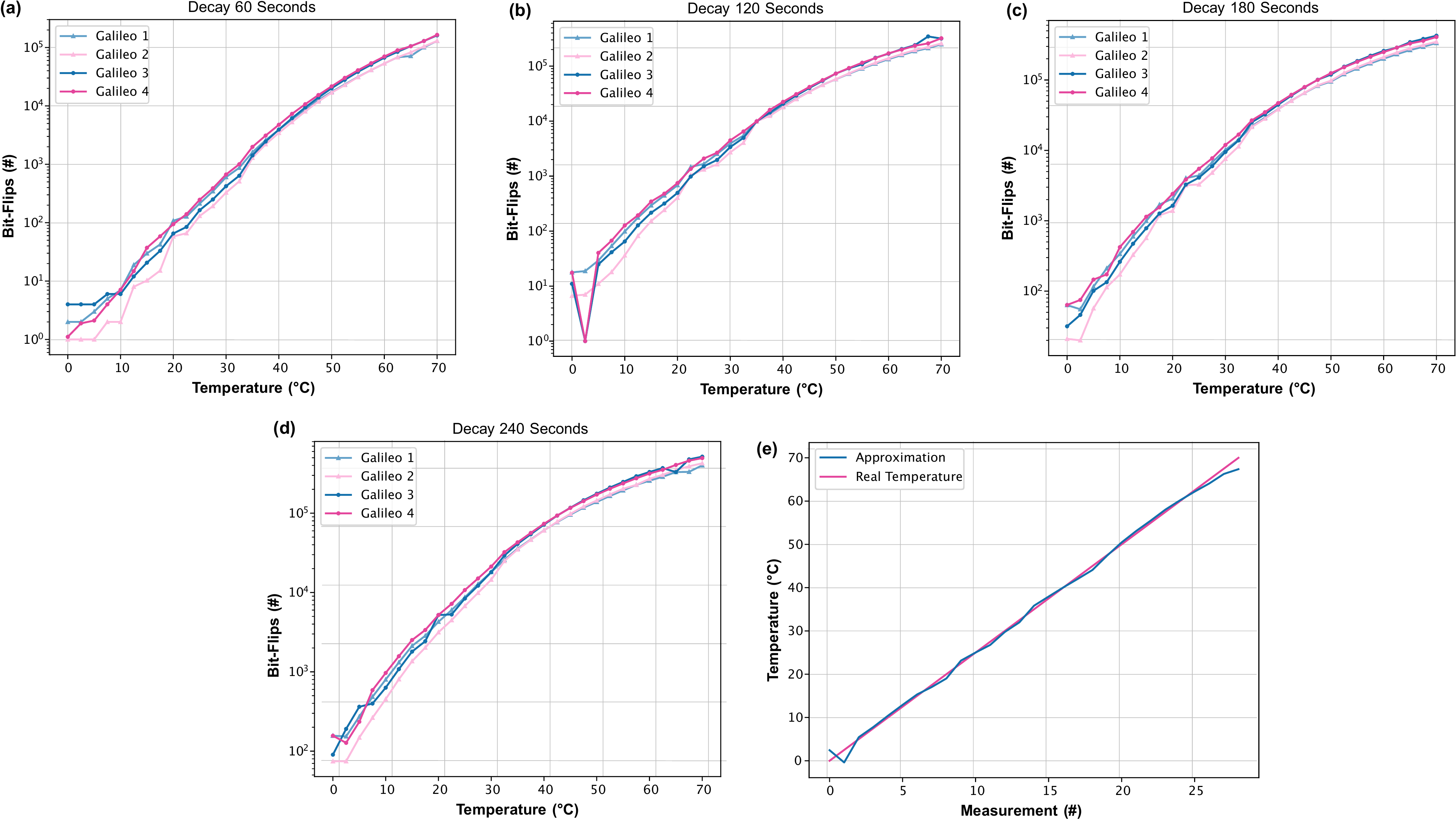}
    \caption{(a) to (d) show the dependency between the temperature and the number of bit-flips of the four decay times \SI{60}{\second}, \SI{120}{\second}, \SI{180}{\second} and \SI{240}{\second}. Each curve shows the average amount of bit-flips of a specific device over ten measurements on the same device. (e) shows the approximation of the temperature given the average bit-flips per temperature of four Galileo boards, based on a decay time of \SI{240}{\second}.}
    \label{fig:approximation}
\end{figure*}

In this section, we introduce a new approximation function that allows us to predict the temperature in the vicinity of a previously unused device by enrollment measurements on a similar device.
As shown in Figure~\ref{fig:approximation}, the dependency between the temperature and the number of bit-flips follows a distribution, which can be approximated by a mathematical function. To achieve a precise approximation, ten measurements are captured for each device, temperature, and decay time. Every curve in Figure \ref{fig:approximation} shows the average amount of bit-flips calculated among these 10 measurements.
In lower temperature regions and with shorter decay times, a higher degree of jittering can be observed. This is caused by the fact that, in lower temperature regions and with smaller decay times, fewer bit-flips occur (for temperatures close to zero, fewer than 10 bit-flips per $1$MiB memory area, even with a decay time of \SI{240}{\second}). Thereby, a small amount of unstable memory cells causes a greater deviation.
In higher temperature regions, the dependency of bit-flips on the temperature gets more precise.
The Galileo boards can only be tested up to a temperature region of \SI{70}{\celsius}, which is the maximum allowed operating temperature for the Intel Quark X1000 SoC~\cite{quark}. When exceeding this temperature, the Galileo boards shut down automatically to avoid hardware damage. Some of the boards shut down with temperature values close to \SI{70}{\celsius}, that's why in temperature regions close to \SI{70}{\celsius} also a more unstable behavior can be observed.

We propose the following function to be used for the approximation of the dependency between the ambient temperature and the number of bit-flips, both for the initial enrollment and to later spy on the temperature:
\begin{equation} 
	T_{apx} = c_1  \cdot e^{c_2 \;  \cdot \; ({\sf bf}^{T}  \cdot \;p)}.
	\label[formula]{formula:approximation_temperature_bitflips}
\end{equation}
Here, a temperature $T_{apx}$ is approximated, given ${\sf bf}^{T}$, the average number of bit-flips of all tested Galileo boards with a decay time of $t=\SI{240}{\second}$ depending on the temperature $T$. The decay time of \SI{240}{\second} allows the best approximation because there, the most bit-flips occur and unstable cells distort the distribution less than in measurements with lower decay times and, thus, fewer bit-flips. The parameter $p$ is calculated as~follows:
\begin{equation}
	p = \frac{{\sf bf}^{T_{k}}_{enr}}{{\sf bf}^{T_k}_{obs}},
\label{equation:p}
\end{equation}
and relates the number of bit flips observed in a particular board under a known temperature $T_{k}$, denoted by ${\sf bf}^{T_{k}}_{obs}$, to the ${\sf bf}_{enr}^{T_{k}}$ that has been observed during the enrollment (in the board used for the enrollment) for temperature $T$, denoted by ${\sf bf}_{enr}^{T}$, if $T_{K}=T$, or the number of bit-flips that is calculated for $T_{K}$ based on the enrollment measurement for $T$ and \Cref{formula:approximation_temperature_bitflips}. Thus, $T_{k}$ can be any temperature as long as ${\sf bf}^{T_{k}}_{obs}$ and ${\sf bf}_{enr}^{T_{k}}$ correspond to devices of the same model, which have the same constants $c_1$ and $c_2$.
Obviously, for the measurements of the same board, $p$ is 1. 
The constants $c_1$ and $c_2$ are optimized during the enrollment process that utilises the whole temperature range, such that $T_{apx}$ is close to $T$, given ${\sf bf}_{enr}^{T}$. Hence, the constants $c_1$ and $c_2$ are calculated based on the following experiment, in which the average number of bit-flips ${\sf bf}^T_{enr}$ of the four Galileo boards is collected for each temperature $T$ during the enrollment:
\[
Exp^{240s} \coloneqq \{{\sf bf}^{\SI{0}{\celsius}}_{enr}, {\sf bf}^{\SI{2.5}{\celsius}}_{enr},..., {\sf bf}^{\SI{70}{\celsius}}_{enr}\}.
\]

The results of this experiment are shown in \Cref{fig:approximation} (d), as the number of bit-flips is approximated for all temperatures in the temperature range from \SI{0}{\celsius} to \SI{70}{\celsius}, using the enrollment measurements of the experiment and \Cref{formula:approximation_temperature_bitflips}. To get a more precise approximation, the constants $c_1$ and $c_2$ are calculated for the temperature regions from \SI{0}{\celsius} to \SI{25}{\celsius}, from \SI{25}{\celsius} to \SI{45}{\celsius}, as well as from \SI{45}{\celsius} to \SI{70}{\celsius}.  In \Cref{fig:approximation} (e), the real temperature is shown in comparison to the approximation function. There, the temperature region is subdivided into the three aforementioned intervals and the approximation is calculated for each interval. For higher temperature values, the function $T_{apx}$ approximates the temperature very precisely. The approximation of values close to zero is more imprecise, due to the small number of bit-flips as described in the previous section.

To spy on the temperature of a previously unseen device (of a known model for which other devices have been fully enrolled), the number of bit-flips ${\sf bf}^{T}_{spy}$ captured at an unknown temperature $T$ is used as ${\sf bf}^{T}$ in \Cref{formula:approximation_temperature_bitflips}. Additionally, the parameter $p$ needs to be calculated using \Cref{equation:p}, and a single enrollment measurement ${\sf bf}^{T_{k}}_{spy}$ of the attacked device under a known temperature $T_{k}$, which acts as the ${\sf bf}^{T_{k}}_{obs}$. In this way, for example, the enrollment measurements on the four Galileo boards can be used to approximate the temperature based on the number of bit-flips, independent of the Galileo board that is used to spy on the temperature. Better results can be achieved at higher temperatures due to more bit-flips and a response with higher stability.

The need for only a single enrollment measurement allows us to execute the attack much more efficiently because a smaller amount of data needs to be transmitted compared to an enrollment that utilises the whole temperature range. Additionally, the extrapolation using $p$ allows a precise temperature approximation on devices with a high deviation in the absolute number of bit-flips, because the dependency between the bit-flips and the temperature will follow the same function in similar devices, notwithstanding the absolute number of bit-flips. Moreover, the extrapolation using $p$, and the fact that devices of the same type demonstrate the same dependency between the number of bit-flips and the temperature, allow for the use of much smaller memory regions, compared to the enrollment measurement that uses the whole temperature range, to spy on the ambient~temperature.

\subsection{Attacks in Practice}
\label{sec:attack_demo}

\begin{figure}[t]
    \centering
    \includegraphics[width=\linewidth]{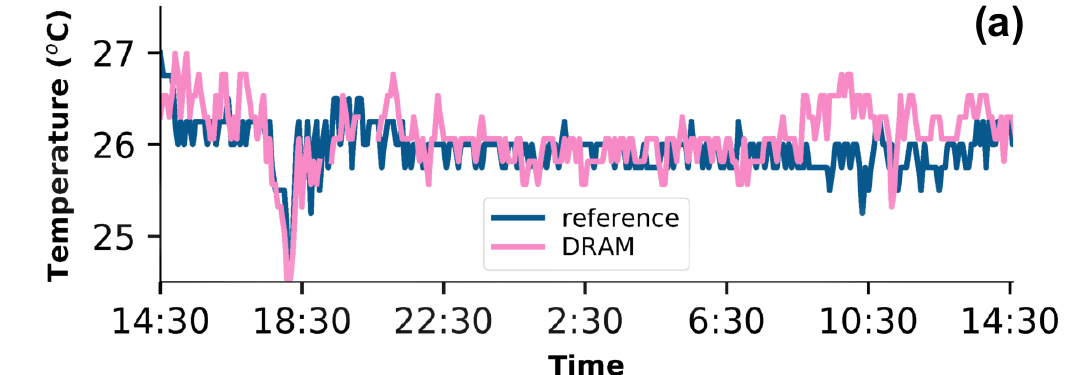}
    \includegraphics[width=\linewidth]{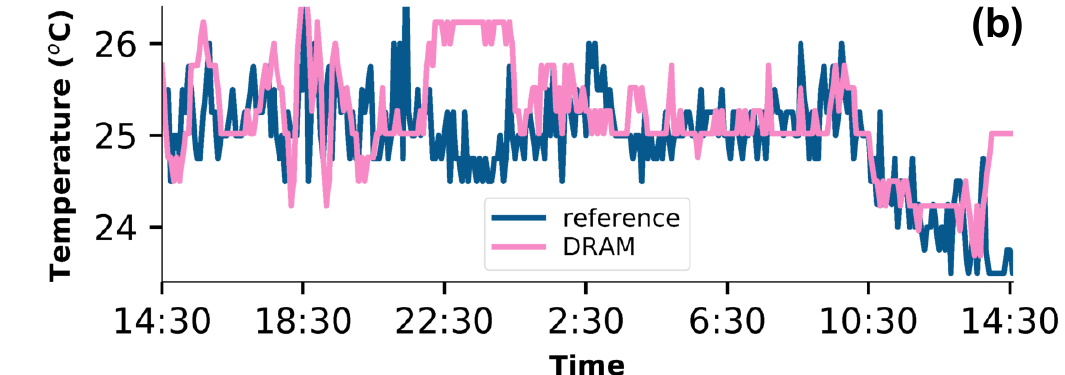}
    \includegraphics[width=\linewidth]{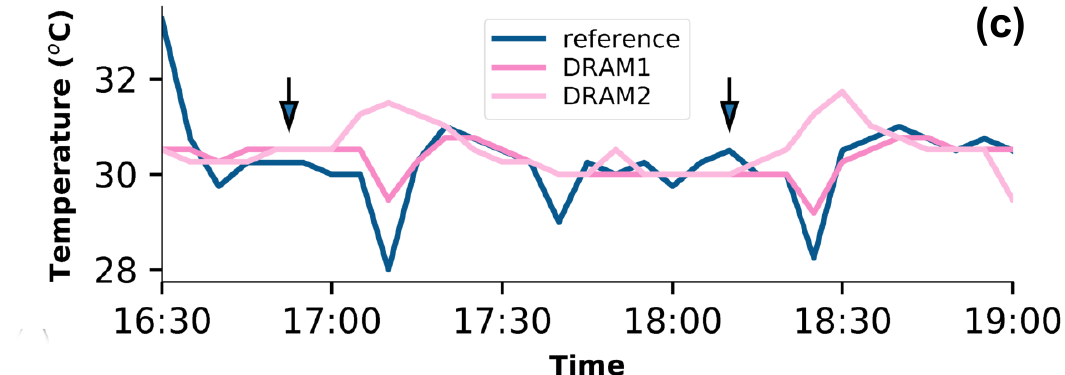}
    \caption{(a), (b) Results of measuring the temperature every \SI{5}{\minute} with DRAM module in two different rooms for 24 hours. (c) Results of measuring the temperature every 5 minutes with DRAM modules in a server rack.
    The arrows show when the server starts to run a job.}
    \label{fig:attack_demo}
\end{figure}

\medskip{\noindent\textbf{Indicator-Cell-Based Attacks.}} This section describes an attack that spies on the temperature in a room, e.g., in a smart home or a data center, based on the evaluation of indicator cells using Intel Galileo boards.
To show the practicability of the attack, we deployed several Intel Galileo boards openly on a table in two rooms and in a server rack. A bare Yocto Linux kernel was used during the test.
We measured the DRAM decay of \SI{60}{\second} with $2$MiB DRAM region every \SI{5}{\minute} to infer the ambient temperature.
For each of the boards, enrollments with decay times ranging from \SI{50}{\second} to \SI{75}{\second} (in steps of \SI{2}{\second}) are taken. According to \Cref{equ:temp} with $k = 0.07$, the enrollments can simulate the ambient temperature change of $[\SI{-3}{\celsius}, \SI{+3}{\celsius}]$. 
Note that, in total, 26 measurements are taken for the temperature range, so the actual temperature resolution is higher than \SI{0.5}{\celsius}.
A thermocouple is used to get the actual temperature during the reference measurements. Indicator cells are generated based on the enrollment measurements and later used to map the decay results to~temperatures. More than twenty candidate indicator cells are found in $2$MiB for each temperature, so $l=21$ is~used.

In Figures~\ref{fig:attack_demo}~(a) and~(b), the temperature in two different rooms measured by the DRAM and a thermocouple for 24 hours is shown. The temperatures measured by the DRAM match the results of the thermocouple. As shown in the figure, during the night, the temperature is stable; while, during the day, due to human activities, the temperature fluctuates in both rooms. Thus, if the attacker can monitor the temperature, this puts the victim's privacy at risk.

In the second experiment, we deployed two Galileo boards in a server rack. \Cref{fig:attack_demo}~(c) shows the temperature measured by two DRAM modules, DRAM1, and DRAM2. DRAM1 is located closer to the fans, with the thermocouple being placed near DRAM1. The arrows indicate when the server starts to run a job for \SI{25}{\minute}. When the server runs, it will gradually heat up DRAM2.
Subsequently, the fan starts working, and ambient temperature drops (especially for DRAM1 and the thermocouple). Consequently, using only the IoT device's DRAM, the attacker can monitor the temperature change of the server, which could create a side-channel to reveal the activity of the~tenants~\cite{islam2017exploiting}.

\medskip{\noindent\textbf{Attack Using the New Approximation Function.}} Here, we consider two different attack scenarios using the approximation function described in Section \ref{sec:approximation_function}. The first scenario concerns spying on the temperature next to a server to gather information about its workload. The second scenario spies the temperature within a room, e.g., in the context of a smart home, to find out if, e.g., somebody is within the room.

These attacks are implemented using two Galileo boards. On the first device, which does not need to be physically in the area being spied, enrollment measurements are taken over the whole temperature range, and these measurements are then used to approximate the dependency of the number of bit-flips on the temperature using \Cref{formula:approximation_temperature_bitflips}. To implement the attacks, the approximation calculated by the measurements of the first device is used as well as a single enrollment measurement of the second Galileo board. This allows an attacker to execute temperature spying attacks on a board without capturing enrollment measurements over the whole temperature range using the spying board itself. 

\begin{figure}[t]
    \centering
    \includegraphics[width=0.8\linewidth]{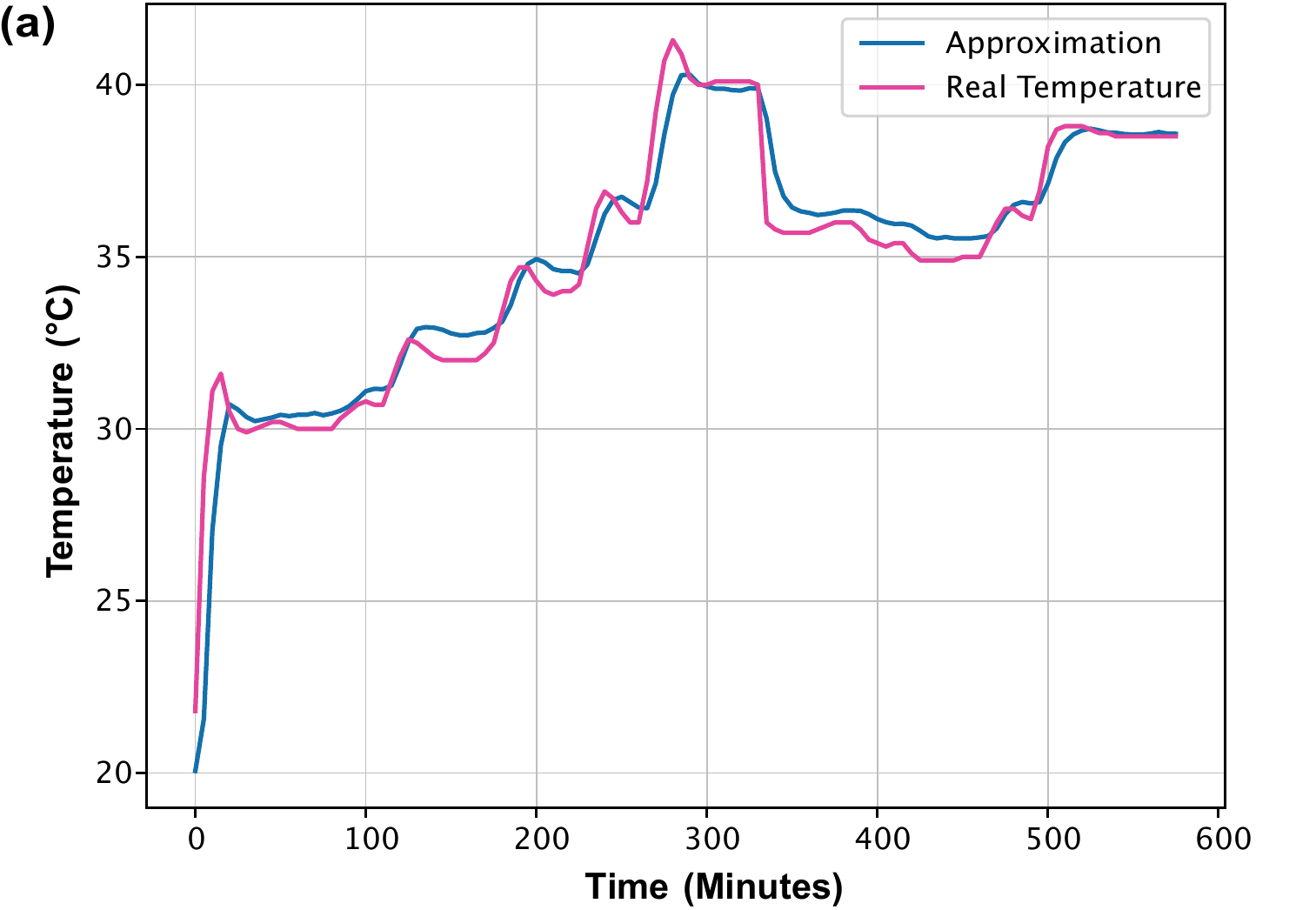}\\[10pt]
    \includegraphics[width=0.8\linewidth]{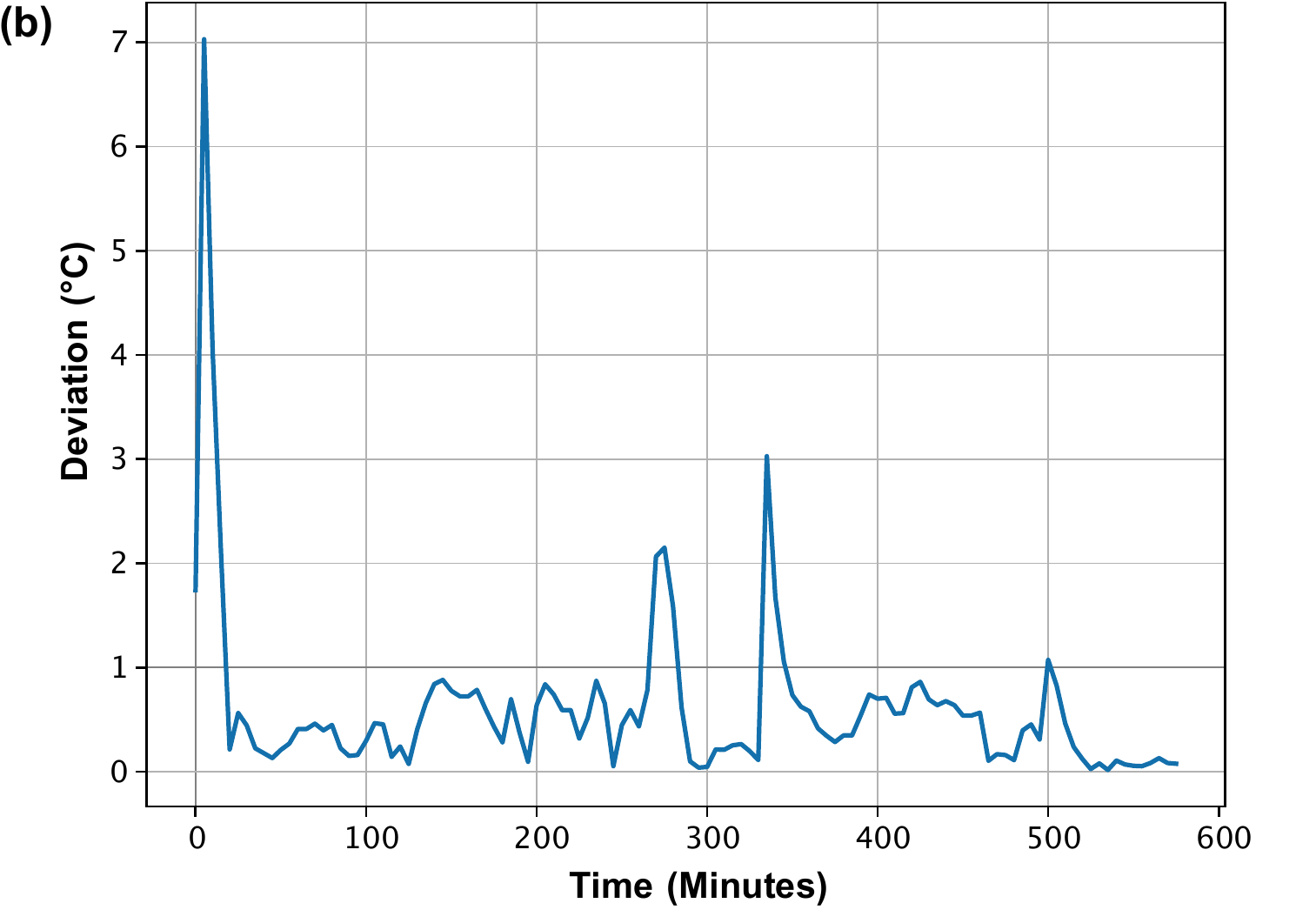}
    \caption{This figure shows a spying attack on the temperature nearby a server system, simulated by a climate chamber. (a) shows a comparison of the real and the approximated temperature using a Galileo board for which only one enrollment measurement has been taken, and \Cref{formula:approximation_temperature_bitflips}. (b) shows the deviation from the real temperature, which is below \SI{1}{\celsius} most of the time.}
    \label{fig:server_sim_temp_precision}
\end{figure}

In the proof-of-concept implementations of both attack scenarios, the spying Intel Galileo board is connected to a local area network via an Ethernet cable. A second malicious device, in our case a Raspberry Pi, is located in a different room, but is connected to the same network.
On the Galileo board, malicious software is installed, which continuously executes a DRAM decay measurement during runtime on a $256$KiBit ($\,$=$~32$KiB) memory region. Before the execution of the measurement, a $256$KiBit memory area of the Galileo board is filled with ones. Afterwards, the memory refresh is disabled, and for each \SI{120}{\second}, the number of bit-flips is collected and sent to the Raspberry Pi. After each measurement, a new \SI{120}{\second} DRAM decay measurement is executed.
In the beginning, only one enrollment measurement is done at \SI{40}{\celsius}. Subsequently, the factor $p$ is calculated according to \Cref{equation:p}. $p$ is stored on the Raspberry Pi and used for each temperature approximation using \Cref{formula:approximation_temperature_bitflips}.
This allows the Raspberry Pi to calculate the temperature in the vicinity of the Galileo board every \SI{128}{\second} when also considering the time to initialize and to read from the memory.

For both scenarios, the ambient temperature was regulated by a Weisstechnik LabEvent climate chamber, which was also controlled by a Raspberry Pi. To test the robustness of the temperature prediction, faster and slower temperature changes were executed. It was also tested how the temperature prediction works on minor temperature variations. 
The temperature curve that was executed, can be seen in \Cref{fig:server_sim_temp_precision}~(a). Here, the approximation using the Galileo board almost follows the real temperature curve. We can see that this approach has a small delay, probably caused by the time needed for the DRAM modules to be affected by the ambient temperature changes. In \Cref{fig:server_sim_temp_precision}~(b), the deviation from the real temperature can be seen. Here, the deviation is below \SI{1}{\celsius} most of the time. Only rapid temperature changes cause greater deviations.

\begin{figure}[t]
    \centering
    \includegraphics[width=0.8\linewidth]{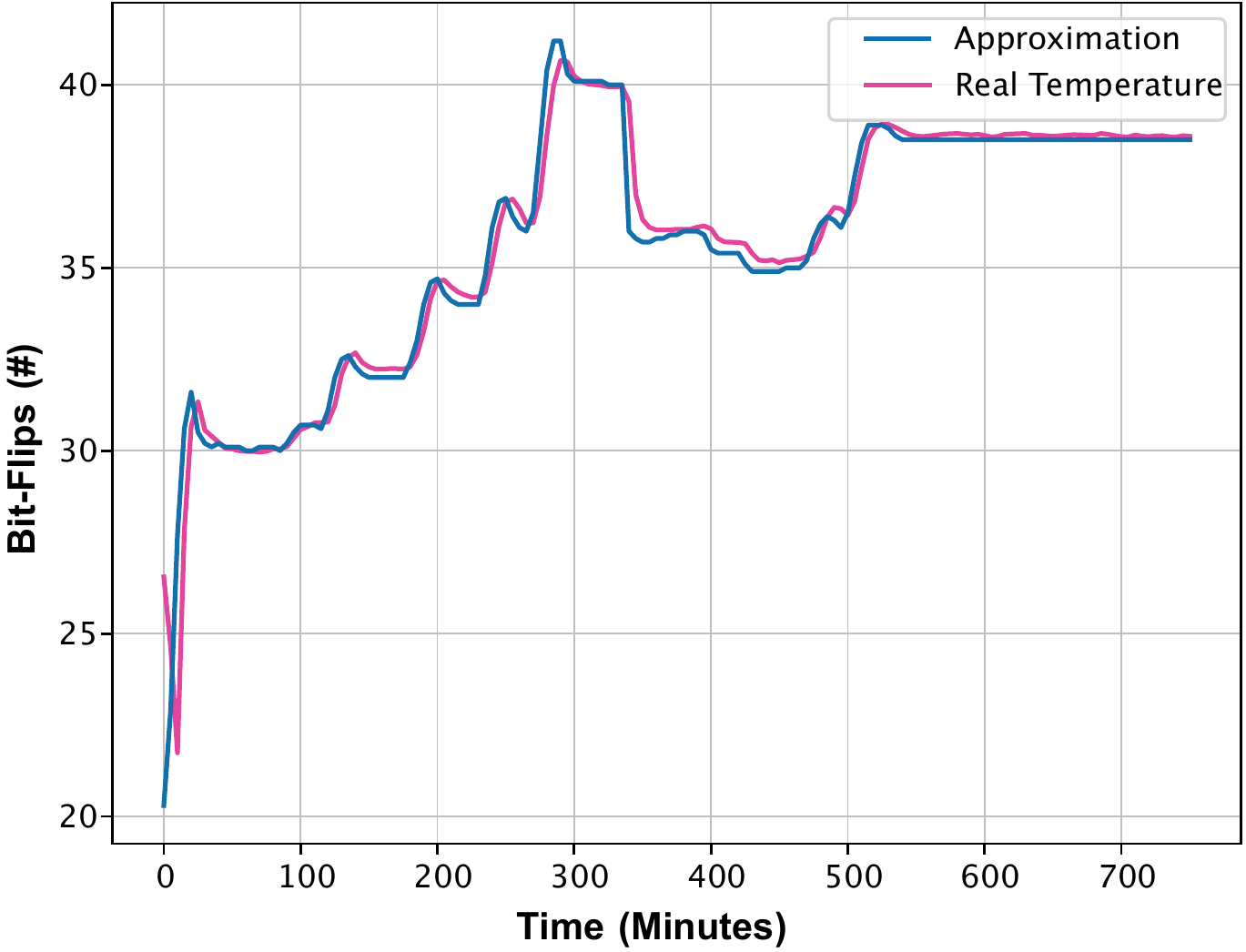}
    \caption{Simulation of a temperature spying attack in the vicinity of a server. In this case, the approximation is based on the device for which an enrollment over the whole temperature range has been performed. As expected, an approximation of the temperature with higher precision is achieved.}
    \label{fig:server_same_dev}
\end{figure}

Afterwards, the same attacks are evaluated on the device previously used for enrollment over the whole temperature range.
As expected, a better approximation of the temperature based on the current DRAM bit-flips of this board can be achieved, in comparison to the execution on the other device, for which an enrollment over the whole temperature range has not been performed. As shown in \Cref{fig:server_same_dev}, the deviation of the approximated and the real temperature is almost always below \SI{0.5}{\celsius}.

\subsection{Attack Complexity}
\label{sec:attack_comp}
The attacker's efforts consist of making enrollment measurements and conducting each temperature readout during the actual attack. For both, the attacker needs to be able to run malicious kernel code on the victim platform to control the DRAM refresh.

The enrollment time consists of the measurement time, the data transfer time, and the time to identify indicator cells. 
The measurement time is the decay time plus the time to initialize (write) and read from the DRAM region.
For a $2$MiB memory region, on an Intel Galileo, it takes about half a minute to read or write the region.
Thus, one enrollment takes about two minutes, considering a decay time of $t=\SI{60}{\second}$: half a minute to initialize, one minute to allow decay to happen and half a minute to read the DRAM region to locate the decayed bits.
By using smaller memory regions of $256$KiBit like during the attacks described in Section \ref{sec:attack_demo}, the additional delay caused by the initialization and the read operations is only \SI{8}{\second}.

The total number of measurements depends on the temperature range and the required temperature resolution. 
To acquire ten enrollment measurements, assuming an average enrollment decay time of $t=\SI{60}{\second}$, it takes less than half an hour.
The data transfer time depends on the size of the data to transfer, and the network speed achieved each time.
The time to compare the enrollment measurement and identify the {indicator cells} is~negligible.

The temperature readout time consists of a single measurement.
Furthermore, because only the indicator cells need to be measured, the time to initialize and read the result is negligible compared to the decay time. 
\section{Countermeasures}  \label{sec:Countermeasures}

In this section, two different countermeasures are discussed to avoid the presented temperature spying attacks.

\subsection{Protection of the Kernel and the Firmware Code}
One way to mitigate the said temperature spying is to prevent disabling the DRAM refresh, 
as the attacker needs to disable the DRAM refresh to measure the DRAM decay.
Since disabling the DRAM refresh can only be achieved in the kernel of the operating system and/or the relevant firmware, on almost all platforms, 
the attacker has to inject untrusted code into the kernel or firmware. 
Thus, one simple countermeasure is to protect the kernel and firmware code. 

However, forcing the DRAM refresh to be always on is not desired from an energy-saving perspective.
To this end, a deep sleep mode usually exists, 
where the DRAM refresh is off. Therefore, an attacker can write initial values into the DRAM region and force the DRAM into that sleep mode, such that the memory decays.
To prevent this attack, the system needs to always zero out the whole memory immediately when the memory wakes up. 

In general, although the implementation of this countermeasure would seem possible, it would also cause a certain level of inconvenience and also potentially require the addition of extra resources to the system. 

\begin{figure}[t]
    \centering
    \includegraphics[width=0.8\linewidth]{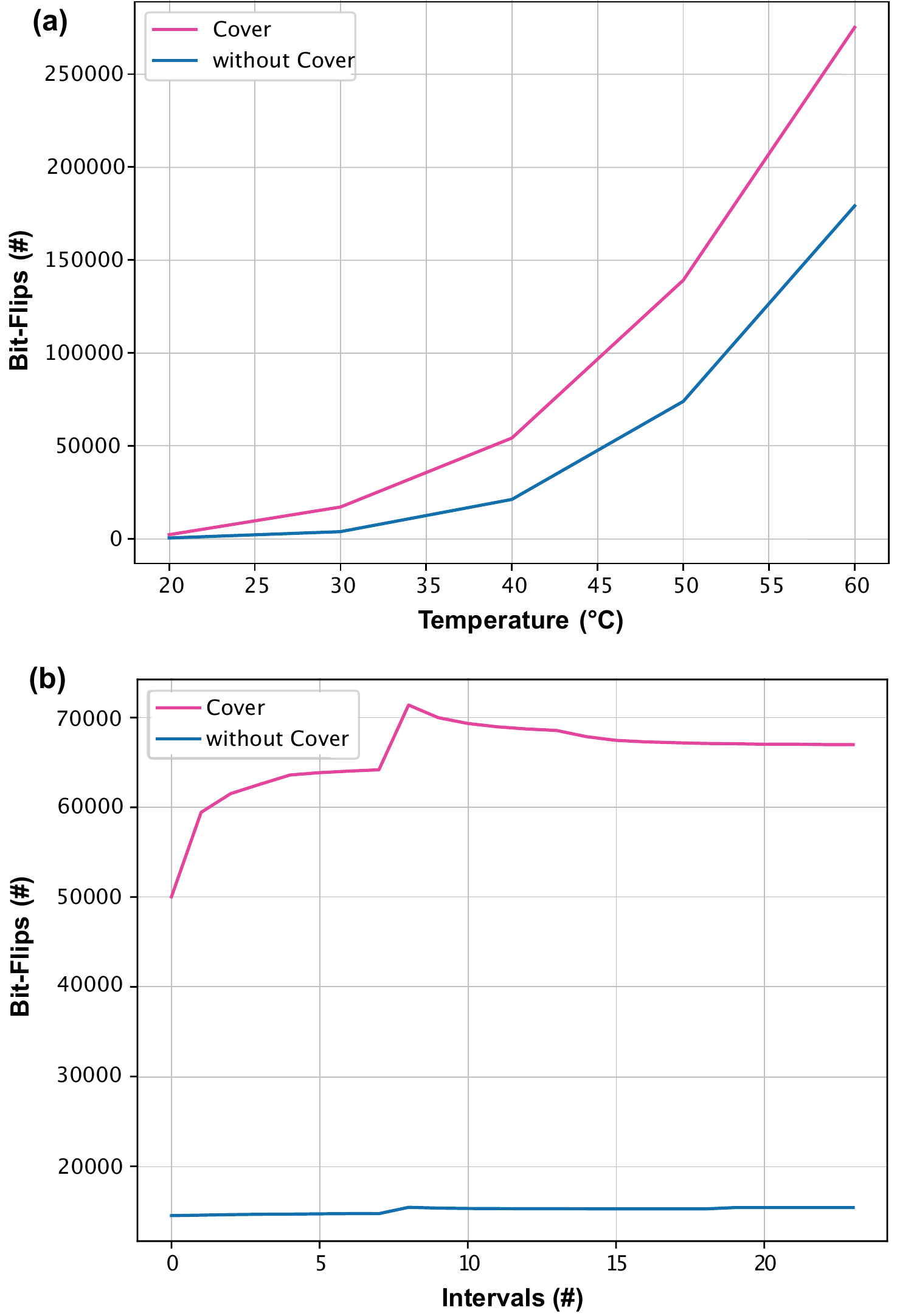}
    \caption{(a) comparison of the number of bit-flips when measuring the boards within a cover box and without it at temperatures from \SI{20}{\celsius} to \SI{60}{\celsius}. Here, the cover causes a higher amount of bit-flips in comparison to the measurement without the cover.
    (b) Execution of 25 measurements with and without the cover at a constant temperature of \SI{40}{\celsius}. The cover causes a much higher amount of bit-flips, compared to the bit-flips caused without it.}
    \label{fig:Cover_20_50_DG}
\end{figure}

\subsection{Influence of Using a Cover}
We also examined the effect of placing the Galileo boards/DRAMs inside a closed box made of PLA (Polylactide). As illustrated in \Cref{fig:Cover_20_50_DG}~(a), more bit-flips occur when the Galileo board is inside the box in comparison to conducting the same experiment without a cover. The observed difference in the number of bit-flips with and without the cover even increases with rising temperature.
For comparison, also the behavior when executing the experiment with constant temperature was examined. 
In \Cref{fig:Cover_20_50_DG}~(b), the number of bit-flips with and without the cover is given in 25 measurements at a constant temperature of \SI{40}{\celsius}. Again, the box causes a much higher amount of bit-flips compared to the execution without the cover. 
Because the cover does not only add a constant offset to the number of bit-flips but also distorts the slope of the function, this mechanism can be used to mitigate temperature spying attacks.
We expect that the higher amount of bit-flips may be caused by the heat produced by the Galileo boards and the missing air circulation.
To make the attack even harder, different covers consisting of diverse materials and varying production parameters, like the thickness of the material, could be used. In effect, the use of a simple box effectively distorts our attacks.

\section{Conclusions}  
\label{sec:Conclusions}
This work demonstrated that commercial, off-the-shelf DRAM modules can be abused to act as remote temperature spy sensors in ordinary IoT devices.
We showed that attackers only need to modify the software of a device and take enrollment measurements at a constant temperature. Subsequently, they can monitor the ambient temperature over a large temperature interval by measuring the DRAM decay while the DRAM refresh operation is disabled. 
We proved in experiments that this approach can achieve a very high temperature resolution of \SI{0.5}{\celsius} in practice.

In addition, this work for the first time suggested and tested countermeasures. The most obvious of these consists of enforcing the DRAM refresh to be turned on permanently. However, this is not desirable from an energy perspective. Another measure is shielding the device against the environment by using a box or a similar encapsulation mechanism. While this can be laborious in practice, it does work effectively, as demonstrated by experiments in this paper. Future research will have to investigate in detail whether there could be other, perhaps yet more effective countermeasures at the circuit or architectural level. The employed attack and analysis code is available under the GPLv3 license at \url{http://caslab.csl.yale.edu/code/tempspy/}.

Our attack once more reminds us that the espionage potential and indirect sensor capacities of electronic IoT devices are currently not well-understood. Seemingly simple components with limited functionalities often can be abused in unforeseen manners; this applies to the recent gyroscope attacks~\cite{michalevsky2014gyrophone} as well as to our novel espionage usage of DRAM cells. In addition, the interplay of several ostensibly trivial system components can often create unforeseen emergent behavior exploitable by attackers. This calls for new, perhaps yet more fundamental research to better protect our security and privacy in relevant environments like the IoT. The first steps towards applying highly developed cryptographic tools like the universal composition framework to complex hardware settings have been made only recently~\cite{canetti2020using}.

\bibliographystyle{IEEEtran}
\bibliography{references}

\begin{IEEEbiography}[{\includegraphics[width=1in,height=1.25in,clip,keepaspectratio]{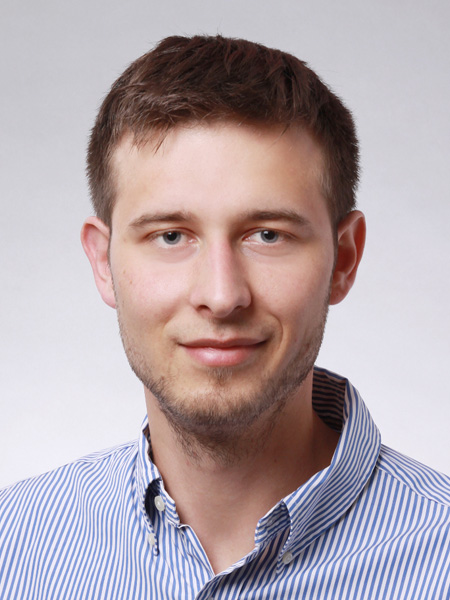}}]{Florian~Frank}
received the B.Sc. degree in computer science from Munich University of Applied Sciences, Germany, in 2018, and the M.Sc. from the University of Passau, Germany, in 2020. 
Currently, he is working towards his Ph.D. degree in the Computer Science Department of the University of Passau. His research interests are hardware security, especially Physical Unclonable Functions, new memory technologies like MRAM, FRAM, and ReRAM, as well as FPGAs. He is a (student) member of the Institute of Electrical and Electronics Engineers (IEEE), and is currently employed as a research assistant in the Chair of Computer Engineering at the University of Passau.
\end{IEEEbiography}

\begin{IEEEbiography}[{\includegraphics[width=1in,height=1.25in,clip,keepaspectratio]{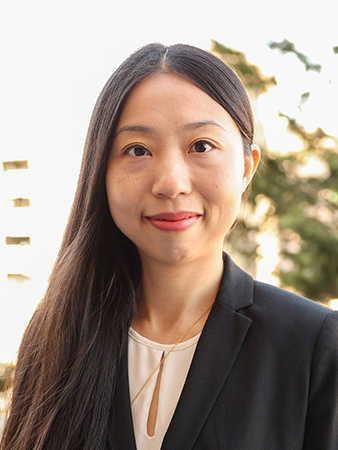}}]{Wenjie~Xiong}
received the B.S. degree in microelectronics and psychology from Peking University, China, in 2014, and the Ph.D. degree from the Department of Electrical Engineering, Yale University, New Haven, Connecticut, in 2020. She is currently an Assistant Professor in the Bradley Department of Electrical and Computer Engineering at Virginia Tech.
Her research interests comprise physically unclonable functions and side-channel attacks and defenses.
\end{IEEEbiography}
\vspace{-20pt}
\begin{IEEEbiography}[{\includegraphics[width=1in,height=1.25in,clip,keepaspectratio]{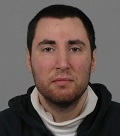}}]{Nikolaos~Athanasios~Anagnostopoulos}
received the B.Sc. degree in computer science from the Aristotle University of Thessaloniki, Greece, in 2012, an M.Sc. degree in computer science from the University of Twente, the Netherlands, and another M.Sc. degree in innovation in information and communication technology from the Technical University of Berlin, Germany, both in 2014. He is currently working towards the Ph.D. degree in the Computer Science Department of the Technical University of Darmstadt, Germany. His research interests include hardware security, with a focus on embedded devices, Physical Unclonable Functions (PUFs), and the Internet of Things (IoT). He is a (student) member of the Institute of Electrical and Electronics Engineers (IEEE), the Association for Computing Machinery (ACM), the German Society for Informatics (Gesellschaft für Informatik -- GI), the International Association for Cryptologic Research (IACR), the Dutch Royal Institute of Engineers (Koninklijk Instituut Van Ingenieurs -- KIVI), the Association of German Engineers (Verein Deutscher Ingenieure -- VDI), the German Association for Electrical, Electronic and Information Technologies (Verband der Elektrotechnik, Elektronik und Informationstechnik -- VDE) and the American Mathematical Society (AMS). 
He is also currently employed as a research assistant in the Chair of Computer Engineering at the University of Passau.
\end{IEEEbiography}
\vspace{-20pt}
\begin{IEEEbiography}[{\includegraphics[width=1in,height=1.25in,clip,keepaspectratio]{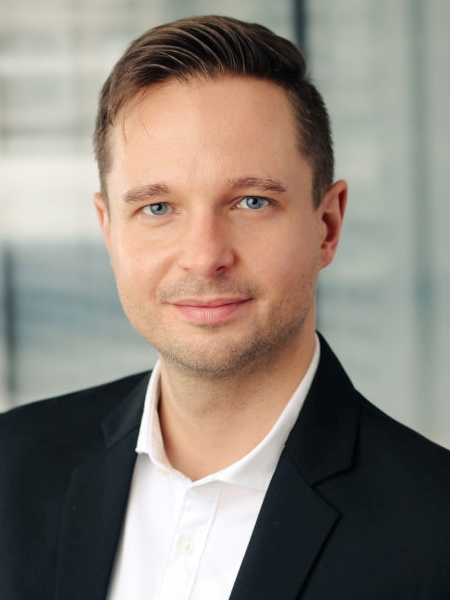}}]{Andr{\'e}~Schaller}
received the M.Sc. degree in computer science in 2012 and the Ph.D. degree in computer science in 2017 from the Technical University of Darmstadt, Germany. He is currently working as an Information Security Engineer at the European Organisation for the Exploitation of Meteorological Satellites (EUMETSAT) as well as a freelance Security Engineer. His main research interests comprise hardware-based cryptography, security of embedded systems, and physically unclonable functions in particular.
\end{IEEEbiography}
\vspace{-20pt}
\begin{IEEEbiography}[{\includegraphics[width=1in,height=1.25in,clip,keepaspectratio]{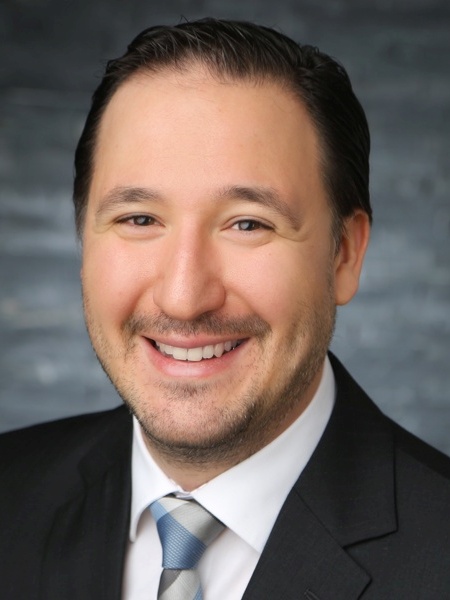}}]{Tolga~Arul}
received the M.Sc. degree in computer science in 2010 and the Ph.D. degree in computer science in 2016 from the Technical University of Darmstadt, Germany.
He has worked as a research associate since 2009 at the Cyber-physical Systems Security Laboratory of the Center for Advanced Security Research Darmstadt and in 2012 joined the National Research Center for Applied Cybersecurity in Darmstadt, Germany.
He is currently a postdoctoral researcher with the Chair of Computer Engineering at the University of Passau, Germany. 
His current research interests include trusted computing and embedded security applied to IoT, transportation, and broadcasting environments. 
He is a member of the Institute of Electrical and Electronics Engineers (IEEE) and the Association for Computing Machinery (ACM).
\end{IEEEbiography}

\begin{IEEEbiography}[{\includegraphics[width=1in,height=1.25in,clip,keepaspectratio]{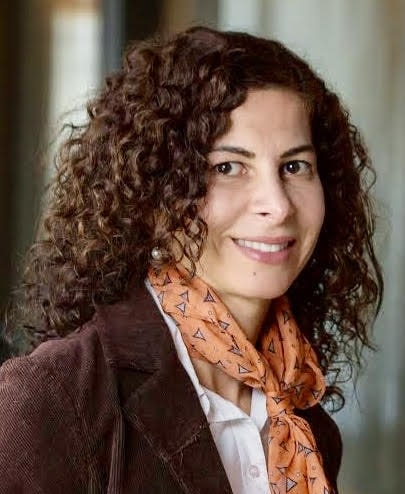}}]{Farinaz~Koushanfar}
is a professor and Henry Booker Faculty Scholar in the Electrical and Computer Engineering (ECE) department at the University of California San Diego (UCSD), where she is the founding co-director of the UCSD Center for Machine Intelligence, Computing \& Security (MICS). She received her Ph.D. in Electrical Engineering and Computer Science as well as her M.A. in Statistics from UC Berkeley. Her research addresses several aspects of efficient computing and embedded systems, with a focus on system and device security, safe AI, privacy preserving computing, as well as real-time/energy-efficient AI under resource constraints, design automation, and reconfigurable computing. She has received several awards and honors for her research, mentorship, teaching, and outreach activities including the Presidential Early Career Award for Scientists and Engineers (PECASE) from President Obama, the ACM SIGDA Outstanding New Faculty Award, Cisco IoT Security Grand Challenge Award, Qualcomm Innovation Award(s), MIT Technology Review TR-35, Young Faculty/CAREER Awards from NSF, DARPA, ONR and ARO, as well as a number of Best Paper Awards. She is a fellow of the IEEE, and a fellow of the Kavli Foundation Frontiers of the National Academy of Sciences.
\end{IEEEbiography}
\vspace{-20pt}
\begin{IEEEbiography}[{\includegraphics[width=1in,height=1.25in,clip,keepaspectratio]{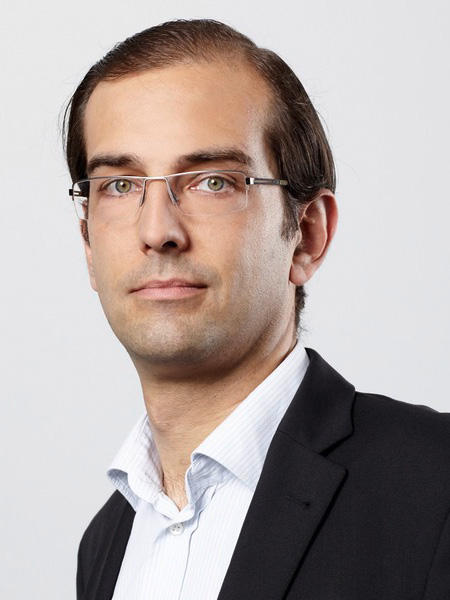}}]{Stefan~Katzenbeisser}
received the Ph.D. degree from the Vienna University of Technology, Austria. After working as a Research Scientist with the Technical University of Munich, Germany, he joined Philips Research as a Senior Scientist in 2006. After holding a professorship for Security Engineering at the Technical University of Darmstadt, he joined University of Passau in 2019, heading the Chair of Computer Engineering. His current research interests include embedded security, data privacy, and cryptographic protocol design. 
\end{IEEEbiography}
\vspace{-20pt}
\begin{IEEEbiography}[{\includegraphics[width=1in,height=1.25in,clip,keepaspectratio]{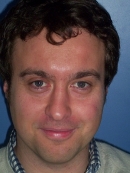}}]{Ulrich~R{\"u}hrmair} 
holds a PhD in computer science from TU Berlin, a PhD in electrical engineering from TU Munich, and an MSc in mathematics from Oxford University. He is a research professor at the University of Connecticut and currently also a guest professor at LMU Munich.  His research interests include applied cryptography and computer security in general, as well as Physical Unclonable Functions (PUFs) and related physical security primitives in particular. He is the founder and current steering committee chair of the ASHES workshop, an annual workshop at ACM CCS since 2017. Furthermore, he is an associate editor at the IEEE Transactions on Information Forensics and Security, Journal of Cryptographic Engineering, Journal on Hardware and Systems Security, and EURASIP Journal on Information Security. Since 2022, he is also a co-speaker of the research focus on {\it “Physics and Security”} at the Center for Advanced Studies at LMU~Munich. 
\end{IEEEbiography}
\vspace{-20pt}
\begin{IEEEbiography}[{\includegraphics[width=1in,height=1.25in,clip,keepaspectratio]{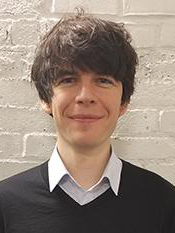}}]{Jakub~Szefer}
received B.S. with highest honors in Electrical and Computer Engineering from University of Illinois at Urbana-Champaign, and M.A. and Ph.D. degrees in Electrical Engineering from Princeton University where he worked with Prof. Ruby B. Lee on secure hardware architectures. He is currently an Associate Professor in the Electrical Engineering department at Yale University, where he leads the Computer Architecture and Security Laboratory (CASLAB). His research interests are at the intersection of computer architecture, hardware security, and FPGA security.
\end{IEEEbiography}
\vfill

\end{document}